\documentclass{elsart}
\usepackage{natbib}
\usepackage{graphicx}
\begin{document}
\runauthor{Cicero, Caesar and Vergil}
\begin{frontmatter}
\title{Tamm states in one dimensional photonic crystals containing left-handed materials}
\author{Abdolrahman Namdar}

\address{Nonlinear Physics Centre, Research School of Physical Sciences and Engineering, Australian National University, Canberra ACT 0200, Australia}
\address{ Physics Department, Azarbaijan University of Tarbiat Moallem, Tabriz, Iran\thanksref{11}}
\thanks[11]{This is permanent address of the author}
\begin{abstract}
We present a theoretical study of  electromagnetic surface waves localized at an interface separating a conventional uniform medium and a semi-infinite
1-D photonic crystal made of alternate left-handed metamaterial and right-handed material  which we refer to as left-handed photonic crystal. We find novel type of surface mode's structure, the so-called surface Tamm states and demonstrate that the
presence of metamaterial in the photonic crystal structure allows for a flexible control of the dispersion properties of surface states, and can support the Tamm states with a
backward energy flow and a vortex-like structure.

\end{abstract}
\end{frontmatter}

\section{Introduction}
The \textit{left-handed metamaterial} (LHM) with negative permittivity and negative permeability has attracted much attention \cite{Smith,Ziolkowski,Shelby} and triggered the debates on the application of the left-handed slab as so-called "superlenses" \cite{Pendry,Garcia}. Over 30 years ago, Veselago \cite{Veselago}, first proposed that this peculiar medium possesses a negative refractive index, which has been demonstrated at microwave frequencies in recent experiment \cite{Shelby}. In such media, there are many interesting properties such as the reversal of both Doppler effect and Chernekov radiation \cite{Veselago}, amplification of evanescent waves \cite{Pendry}, and unusual photon tuneling \cite{Zhang}, negative giant Goos-Hanchen effect \cite{Ilya03}. All these phenomena are rooted in the fact that the phase velocity of light wave in the LHM is opposite to the velocity of energy flow, that is, the Poynting vector and wavevector are antiparallel so that the wavevector, the electric field, and the magnetic field form a left-handed (LH) system. 

Interfaces between different physical media can support a special type of localized waves as \textit{Surface waves} or \textit{surface modes} where the wave vector becomes complex causing the wave to exponentially decay away from the surface. Aside from their intrinsic interest, surface electromagnetics waves have recently been proposed \cite{Kramper,Moreno} as a way to efficiently inject light into a photonic crystal waveguide, or to extract a focused beam from a channel. In periodic systems, staggered surface modes are often referred to as \textit{Tamm states} \cite{Tamm,Shockly} first identified as localized electronic states at the edge of a truncated periodic potential. An optical analog of linear Tamm states has been described theoretically and demonstrated experimentally for an interface separating periodic and homogeneous dielectric media \cite{Yeh1,Yeh2} . In optics, the periodic structures have to be manufactured artificially in order to manipulate dispersion properties of light in a similar way as the properties of electrons are controlled in crystals.   

One-dimensional photonic crystal consisting of alternate LHM and conventional right-handed material (RHM) layers have already been investigated through calculating the transmittance or the reflectance the structures \cite{Zhang,Gerardin}and demonstrating of complete band gap \cite{Ilya2}. In this letter we demonstrate another example of unusual properties of LHM and study electromagnetic surface TE waves guided by an interface between RHM and a 1D periodic structure consisting of alternate LHM and RHM layers which we refer to \textit{left-handed photonic crystal}. We demonstrate that the presence of a LHM allows for a flexible control of the dispersion properties of surface states. 

We pay special attention to the terminating layer (or cap layer) of the periodic structure and assume that it has the width different from the width of other layers of the structure. We study the effect of the width of this termination layer on surface states, and explore a possibility to control the dispersion properties of surface waves by adjusting termination layer thickness. We find \textit{novel types of surface Tamm states} at the interface with the LH photonic crystal. In addition to absolutely novel structure due to sharp jumps at the interface of layers resulting from opposite signs of permeability of adjacent layers, these modes have a backward flow and a vortex-like structure. The surface modes in this case we call \textit{left-handed Tamm states}.
\typeout{SET RUN AUTHOR to \@runauthor}

\section{Transfer matrix method and surface waves}

\begin{figure}
\centering
\includegraphics[width=85mm]{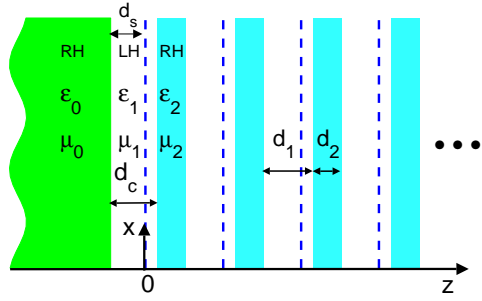}
\caption{Geometry of problem. In our calculations we take the following values: $d_{1}=2.5$cm, $d_{2}=1$cm, $\varepsilon_{0}=1$, $\mu_{0}=1$, $\varepsilon_{1}=-4$, $\mu_{1}=-1$, $\varepsilon_{2}=2.25$, $\mu_{2}=1$}
\label{Fig1}
\end{figure}

We use the transfer matrix method to describe surface Tamm states that form at the interface between an uniform medium and a semi-infinite one dimensional photonic crystal containing negative refraction metamaterial. Geometry of our problem is sketched in Fig.~\ref{Fig1}.
 We choose a coordinate system in which the layers have normal vector along OZ. As shown in Fig.~\ref{Fig1} the crystal is capped by a layer of the same material but different width, $d_{c}$. For convenience of presentation, we split this termination layer of index of refraction $n_{1}$ and width $d_{c}$ into sublayers, of lengths $d_{s}+d_{t}=d_{c}$. The first sublayers extends from $z=-d_{s}$ to $z=0$. Then the periodic array that forms the one dimensional photonic crystal consists of cells each made of three uniform layers of widths $d_{t}$, $d_{2}$, and $d_{1}-d_{t}$ whose respective indices of refraction are $n_{1}, n_{2}$, and $n_{1}$. We consider the propagation of TE-polarized waves described by one component of the electric field, $E$=$E_{y}$, and governed by a scalar Helmholtz-type equation. We look for stationary solutions propagating along the interface with the characteristic dependence, $\sim$ exp[-$i\omega(t-\beta x/c)]$, where $\omega$ is an angular frequency, $\beta$ is the normalized wavenumber component along the interface, and $c$ is the speed of light, and present this equation in the form,
\begin{equation}
\left[ \frac{d^{2}}{dz^{2}}+k_{x}^{2}+\frac{\omega^{2}}{c^{2}}\varepsilon(z)\mu(z)-\frac{1}{\mu(z)}\frac{d\mu}{dz}\frac{d}{dz}\right] E=0
\end{equation}
where $k_{x}^{2}$=$\omega^{2}\beta^{2}/c^{2}$, both permittivity $\varepsilon(z)$ and permeability $\mu(z)$ characterize the transverse structure of the media. Matching the solutions  and derivatives at the two interfaces, one finds the elements of transfer matrix, $M$.
\begin{equation}
M_{11}=e^{ik_{1}d_{1}}\left( cosk_{2}d_{2}+\frac{i}{2}\left( \frac{\kappa_{1}}{k_{2}}+\frac{k_{2}}{\kappa_{1}}\right) {sink_{2}d_{2}}\right) 
\end{equation}
\begin{equation}
M_{12}=e^{ik_{1}(d_{1}-2d_{t})} \frac{i}{2}\left(\frac{\kappa_{1}}{k_{2}}+\frac{k_{2}}{\kappa_{1}}\right) {sink_{2}d_{2}}
\end{equation}

 \begin{equation}
M_{21}=-e^{-ik_{1}(d_{1}-2d_{t})} \frac{i}{2}\left(\frac{\kappa_{1}}{k_{2}}+\frac{k_{2}}{\kappa_{1}}\right) {sink_{2}d_{2}}
\end{equation}

\begin{equation}
M_{22}=e^{-ik_{1}d_{1}}\left( cosk_{2}d_{2}-\frac{i}{2}\left( \frac{\kappa_{1}}{k_{2}}+\frac{k_{2}}{\kappa_{1}}\right) {sink_{2}d_{2}}\right) 
\end{equation}
where $k_{i}=k\sqrt{n_{i}^{2}-\beta^{2}}$, $i=1,2$ and $\kappa_{1}=\frac{\mu_{2}}{\mu_{1}}k_{1}$. The parameters $\mu_{1}$ and $\mu_{2}$ are permeability of layers in photonic crystal.

 Surface modes correspond to localized solutions with the field $E$  decaying from the interface in both the directions. In the left-side homogeneous medium ($z<-d_{s}$, see Fig.~\ref{Fig1}), the fields are decaying provided $\beta>\varepsilon_{0}\mu_{0}$ . In the right side periodic structure, the waves are the Bloch modes
\begin{equation}
E(z)=\Psi(z)exp(iK_{b}z)
\end{equation}
where $K_{b}$ is the Bloch wavenumber, and $\Psi(z)$ is the Bloch function which is periodic with the period of the photonic structure (see details, e.g., in Ref. \cite{Yeh3}). In the periodic structure the waves will be decaying provided $K_{b}$ is complex; and this condition defines the spectral gaps of an infinite photonic crystal. For the calculation of the Bloch modes, we use the well-known transfer matrix method \cite{Yeh3}. 

To find the Tamm states, we take solutions of Eq. (1) in a homogeneous medium and the Bloch modes in the periodic structure and satisfy the conditions of continuity of the tangential components of the electric and magnetic fields at the interface between homogeneous medium and periodic structure~\cite{Martorell}.
\section{Results and discussion}
\begin{figure}
\centering
\includegraphics[width=85mm]{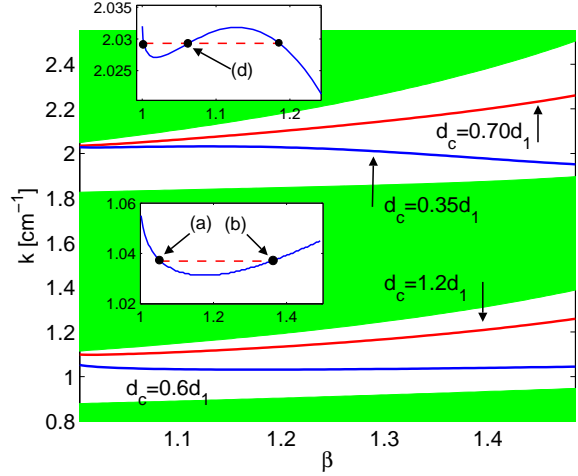}
\caption{Dispertion properties of the Tamm states in the first and second spectral gaps. Shaded: bands of one-dimensional LH photonic crystal. Solid: dispersion of the Tamm states with LHM. Corresponding values of the cap layer thickness are indicated next to the curves, Lower and upper insets show blow-up regions of small $\beta$ for $d_{c}=0.6d_{1}$ and $d_{c}=0.35d_{1}$ respectively. Points (a), (b), and (d) correspond to the mode profiles presented in Figs.~\ref{Fig3}(a, b, d).}
\label{Fig2}
\end{figure}
We summarize the dispersion properties of the Tamm states in the first and second spectral gaps on the plane of the free-space wavenumber $k=\omega/c$ versus the propagation constant $\beta$ (see Fig. \ref{Fig2})
for different values of the cap layer thickness $d_{c}$.

As mentioned above, Tamm states exist in the gaps of the photonic bangap spectrum (unshaded regions in Fig. \ref{Fig2}). An important point that we can see from Fig. \ref{Fig2} is that the sign of the slope of the dispersion curve changes at one (or two) extremum point at first (or second) bandgap, respectively. The slope of the dispersion curve determine the corresponding group velocity of the mode. According to the energy flow of modes (see Fig. \ref{Fig4}) positive (negative) slope of dispersion curve, here, correspond with backward (forward) energy of mode which we call backward (forward) Tamm states. So, for a thin cap, $d_{c}=0.6d_{1} $, the backward Tamm states occur at the incidence angles corresponding $\beta>1.17$, while backward Tamm state for thick cap $d_{c}=1.2d_{1}$ take place at wide ranges $\beta>1.02$. Meanwhile the LH Tamm surface wave has a \textit{vortex-like energy flow pattern}.

As a result of different slope of the dispersion curve of the LH Tamm states, we observe the mode degeneracy, i.e, for the same frequency $\omega$ (or wavenumber $k$), there exist two modes with different value of $\beta$ in first bandgap (see lower inset in Fig.~\ref{Fig2}) while, we can see three modes with different value of $\beta$ in second bandgap (see upper inset in Fig.~\ref{Fig2}). There are modes with negative and positive group velocities (with respect to the propagation wavevector). Such modes are termed as \textit{forward} and \textit{backward} , respectively. In the forward wave, the direction of the total energy flow coincides with the propagation direction, while in the backward wave the energy flow is backwards with respect to the wavevector. Physically, this difference can be explained by looking at the transverse structure of modes.\begin{figure}
\centering
\includegraphics[width=85mm]{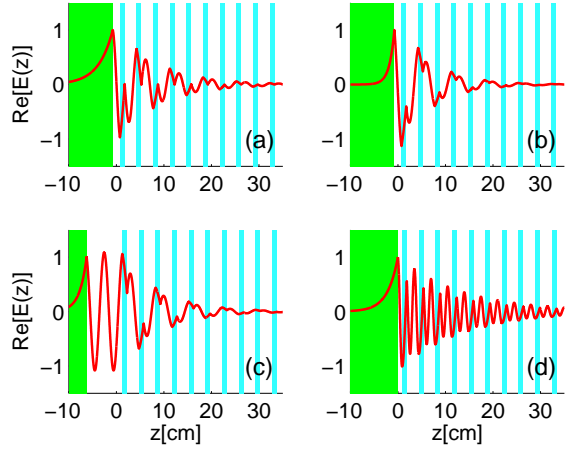}
\caption{Examples of the LH surface Tamm states. (a) Forward LH mode, $d_{c}=0.6d_{1}$, $k=1.037$cm$^{-1}$, $\beta=1.05$; (b) Backward LH mode, $d_{c}=0.6d_{1}$, $k=1.037$cm$^{-1}$, $\beta=1.36$; (c) Guided backward LH mode, $d_{c}=3d_{1}$, $k=0.9965$cm$^{-1}$, $\beta=1.2$; (d) Backward LH second bandgap mode, $d_{c}=0.35d_{1}$, $k=2.029$cm$^{-1}$, $\beta=1.?$;}   
\label{Fig3}
\end{figure} In Figs. ~\ref{Fig3} (a,b)
we plot the profiles of the two modes having the same frequency $k=\omega/c=1.037cm^{-1} $, with different longitudinal wavenumber $\beta$. Corresponding points are shown in the lower inset in Fig.~\ref{Fig2}. For the mode (a), the energy flow in the homogeneous RHM exceeds that effective backward energy flow in the periodic structure (slow decay of field for $z<-d_{s}$ and fast decay into the periodic structure), thus the total energy flow is forward. For the mode (b) we have the opposite case, and the mode is backward. To demonstrate this, in Fig.~\ref{Fig4} 
we plot the total energy flow in the modes as a function of the wavenumber $\beta$ for first and second bandgap surface modes. These results confirm our discussion based on the analysis of the dispersion characteristics. One can see from Fig.~\ref{Fig4} that the increasing the cap layer thickness, $d_{c}$ leads to wide range of backward modes (campare energy flow of $d_{c}=0.6d_{1}$ and $d_{c}=1.2d_{1}$). In the conventional geometry consisting of RHM layers hence right-handed (RH)Tamm state the energy flow in all parts of the wave are directed along the wave number and the localized surface waves are always forward.
\begin{figure}
\centering
\includegraphics[width=85mm]{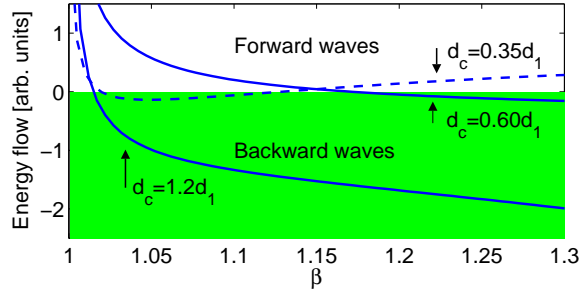}
\caption{Total energy flow in first (solid) and second (dashed) bandgap Tamm modes vs. $\beta$. Corresponding values of the cap layer thickness are indicated next to the curves.}
\label{Fig4}
\end{figure}
An example of the second-gap mode profile is shown in Fig.~\ref{Fig3}(d). One can see that the second band modes are generally weaker localized at the interface then the modes from the first band gap. In addition, the oscillation period of the transverse structure of modes at the first bandgap is twice of the physical structure of photonic crystal while these are equal at the second bandgap.

\begin{figure}
\centering
\includegraphics[width=85mm]{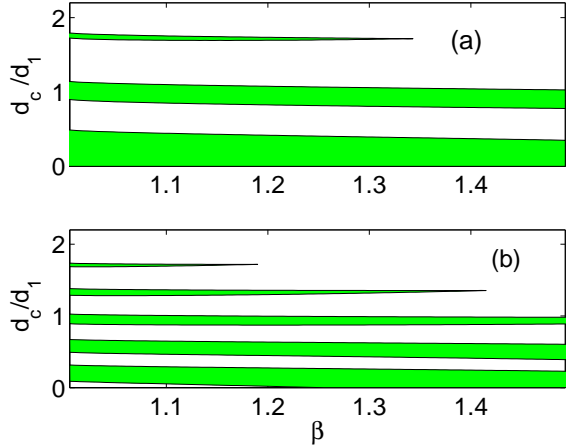}
\caption{Existence regions of the surface Tamm modes for first bandgap (a) and second bandgap (b). The modes do not exist in the shaded regions.}
\label{Fig5}
\end{figure}

Finally, in Fig.~\ref{Fig5} we plot the existence regions for the surface Tamm modes for the both first (Fig.~\ref{Fig5}(a))  and second (Fig.~\ref{Fig5}(b)) bandgaps on the parameter plane ($d_{c},\beta$). Shaded is the area where the surface modes do not exist. Increasing the cap layer thickness, $d_{c}$, we effectively obtain a dielectric waveguide, one cladding of which is a homogeneous medium, while the other one is a LH photonic crystal. In such a case a typical mode is shown in Fig.~\ref{Fig3}(c). Fig.~\ref{Fig5} shows that the non-existence regions of Tamm states qualitatively different for first and second bandgaps. While there exist three discrate non-existence regions of Tamm states for first bandgap, we found five discrate non-existence regions of Tamm states for second bandgap including backward and forward modes.
\section{Conclusions}
We have studied novel types of electromagnetic surface waves guided by an interface between a homogeneous conventional material and a one-dimensional LH photonic crystal consisting of alternate LHM and RHM layers. We have shown that in the presence of a LHM in the photonic crystal structure the surface Tamm waves firstly have a absolutely novel shape with a sharp jumps in the interface of layers due to opposite signs of the permeability in two adjacent media, secondly can be either forward or backward. It must be noted that Tamm states for conventional structure are always forward. We have studied the properties of the first and second bandgap Tamm states and analysed the existence regions for both band modes. We believe our results will be useful for a deeper understanding of the properties of surface waves in plasmonic and metamaterial systems.

\textbf{Acknowledgments}

The author thanks to Prof. Yuri S. Kivshar and Dr. Ilya V. Shadrivov for valuable discussions and are particularly grateful to the former for a careful reading of the manuscript, which clarified a number of important points. This work was supported by the Azarbaijan University of Tarbiat Moallem.


\end{document}